# Direct Observation of One-Dimensional Peierls-Type Charge Density Wave in Twin Boundaries of Monolayer MoTe$_2$

Li Wang[†,⊥], Ying Wu[†,⊥], Yayun Yu[‡,⊥], Aixi Chen[†], Huifang Li[†], Wei Ren[†,∥], Shuai Lu[†], Sunan Ding[*,†,∥], Hui Yang[†], Qi-Kun Xue[§], Fang-Sen Li[*,†,∥], Guang Wang[*,‡,§]

[†] Vacuum Interconnected Nanotech Workstation (Nano-X), Suzhou Institute of Nano-Tech and Nano-Bionics (SINANO), Chinese Academy of Sciences (CAS), Suzhou 215123, China

[‡] Department of Physics, College of Liberal Arts and Sciences, National University of Defense Technology, Changsha 410073, China

[§] State Key Laboratory of Low-Dimensional Quantum Physics, Department of Physics, Tsinghua University, Beijing 100084, China

[∥] School of Nano-Tech and Nano-Bionics, University of Science and Technology of China, Hefei, 230026, China

**Abstract**

One-dimensional (1D) metallic mirror-twin boundaries (MTBs) in monolayer transition metal dichalcogenides (TMDCs) exhibit a periodic charge modulation and provide an ideal platform for exploring collective electron behavior in the confined system. The underlying mechanism of the charge modulation and how the electrons travel in 1D structures remain controversial. Here, for the first time, we observed atomic-scale structures of the charge distribution within one period in MTB of monolayer MoTe$_2$ by using scanning tunneling microscopy/spectroscopy (STM/STS). The coexisting apparent periodic lattice distortions and U-shaped energy gap clearly demonstrate a Peierls-type charge density wave (CDW). Equidistant quantized energy levels with varied periodicity are further discovered outside the CDW gap along the metallic MTB. Density functional theory (DFT) calculations are in good agreement with the gapped electronic structures and reveal they originate mainly from Mo 4d orbital. Our work presents hallmark evidence of the 1D Peierls-type CDW on the metallic MTBs and offers opportunities to study the underlying physics of 1D charge modulation.

Exotic collective excitation behaviors of electrons in 1D systems, including charge density modulation at the periodicity of lattice distortion in metallic chains,[1-3] Majorana bound states in semiconductor nanowires,[4] and topological edge states in higher-order topological insulator,[5] provide promising platforms to investigate the quantum physics and novel applications. In a 1D metallic state, the charges modulate periodically below CDW transition temperature ($T_{CDW}$) due to the Peierls instability.[1] The folded Brillion-zone in momentum space opens an electron bandgap at the Fermi wave vector ($k_F$), which overcompensates the Coulomb and elastic energy cost due to periodic lattice distortions. However, how electrons are confined in 1D structures remains far from clear, which results from the strong interactions between quasi-1D atomic chains,[3] nanowires,[2] superconductors,[6] and their supporting substrates.

Recently, the conductive MTB embedded inside monolayer TMDC has been recognized as a newly discovered 1D metallic system.[7] The underlying mechanism of the charge modulation and electronic properties of different 1D MTBs are controversial,[8-12] even with similar experimental phenomena, such as electron density modulation with a 3$a$ period in a so-called 4|4P MTB [9,10,12] and 2$a$ period in so-called 4|4E MTB,[11] suppression of the density of states (DOS) near the Fermi energy ($E_F$) at low temperature, and enhanced peaks cross $E_F$. Barja *et al.* [10] attributed the charge density modulation in MoSe$_2$/bilayer graphene/SiC(0001) and an energy gap to the formation of CDW by theoretically considering lattice distortion. Liu *et al.* [9] argued that the interaction between MoSe$_2$ and highly ordered pyrolytic graphite substrate forms a quantum well barrier, which induces different periodicity of electron density modulation at corresponding mapping energies. By focusing on 4|4E MTB and taking proper electron-electron interaction into account, Jolie *et al.* [11] rule out the CDW state in MoS$_2$/graphene/Ir(111) system, but report a Tomonaga-Luttinger liquid (TLL) ground state with charge-spin separation, according to the observed non-equidistant quantized satellite peaks and unusual spatial distributions along with the MTB. Most recently, on 4|4P MTB in the MoSe$_2$/graphene system,[12] a signal of TLL state was claimed to be present. The rich emergent phenomena [7-16] in 1D MTBs imply that the electron-lattice and electron-electron interactions play important roles in the 1D metallic systems. However, the atomic-scale imaging of the detailed structure within one charge modulation period has not been directly observed yet. Therefore, deterministic experimental evidence is necessary to distinguish CDW or TLL ground state in the typical 1D metallic system.

We have systematically studied 1D CDW states in MTB of monolayer MoTe$_2$ and investigated atomic-scale CDW order, electronic band structure, and spatial conductance distribution by using STM/STS at liquid nitrogen temperature (77K). An electron density modulation with 3$a$ periodicity is observed together with an asymmetric U-shaped energy gap opened across $E_F$. Constant-height conductance mapping resolves all three spots within one 3$a$ period, which exhibits an apparent lattice displacement. We have further discovered the equidistant quantized peaks outside the energy gap, as well as varied periodicity along the MTB. The formation of CDW order is consistent with our DFT calculations based on Peierls distortion, which demonstrates that the gapped electronic band structures mainly originate from Mo 4d orbital.

**RESULTS AND DISCUSSION**

Monolayer 2H-MoTe$_2$ is epitaxially grown on bilayer graphene (BLG) terminated SiC(0001) substrate by ultra-high vacuum (UHV) molecular beam epitaxy (MBE) [17,18] (see the Methods). The morphology and electronic structure of a typical MTB of monolayer MoTe$_2$ films are shown in Figure 1. A network of domain boundaries with triangular shape (also see Figure S1a) is always observed due to translational lattice shifts between various nucleation centers during MBE growth,[8-12,19-21] which are identified as MTBs with Te atom bonding with four Mo atoms (see the ball model in Figure 1a). Along the domain boundary, we observe two parallel rows of big bright spots. Similar results have been previously observed in MoSe$_2$ [10] and MoTe$_2$ [14] films *via* non-contact atomic force microscopy [10] and scanning transmission electron microscopy.[14] The triangle domain confines the electron within well-defined 1D boundaries. Independent of length, we find that the spots are separated with the same modulation period of ~ 3$a$ along the MTB ($a$ = 3.52 Å is the lattice constant of MoTe$_2$). Such charge modulation is much more apparent when performing the normal constant-height STS mapping (see Figure 1b). The mappings at different sample biases yield the same modulation periodicity (as shown in Figure S1e), similar to the case [10] of MoSe$_2$ on BLG/SiC(0001), indicating the absence of quantum well states in the 1D MTB of MoTe$_2$ here.

We have conducted tunneling spectroscopy measurements to probe the local density of states (LDOS). Away from MTB, monolayer MoTe$_2$ exhibits no moiré pattern but a (1×1) atomic hexagonal lattice (also see Figure S1a), suggesting a rather weak interaction between MoTe$_2$ and bilayer graphene. The d$I$/d$V$ spectra on the (1×1) structure show a large bandgap of 2.01 eV in Figure 1c, in good agreement with our previous measurements [17] and other works.[10, 14] The d$I$/d$V$ tunneling spectra on the MTB show a little narrowed gap of 1.80 eV. We notice that there are some non-negligible metallic states within the gap. The inset of Figure 1c shows various typical features. Two asymmetric coherence peaks (marked as φ+ and φ-) form a U-shaped energy gap of 152 meV around the Femi level. The electron-hole asymmetry may be induced by lattice strain or scattering from edges of finite 1D length.[22] We find that the gap size changes from one to another MTB, varying from 288 meV to 90 meV, but remains constant along the same MTB. Perpendicular to the MTB, the spatially resolved d$I$/d$V$ tunneling spectra (Figure 1d) are uniform, not only in the size of the opened gap but also the attendant satellite peaks (also see the raw data of the line spectroscopy in Figure S2). It illustrates the 1D nature of MTBs.

To achieve high-resolution atomic imaging within the 3$a$ modulation period, we have performed the constant-height tunneling d$I$/d$V$ mapping (see the detail in Supporting Information). We mapped the spatial distribution at electron band-edge states (φ+ and φ- in Figure 1c). Figure 2a shows the mapping image at energy of 80 meV (φ+) for MTB-1 and MTB-2 marked in Figure 1b. Rather than big bright spots in Figure 1b, we observe well-resolved structures: type A, pairs of two bright spots (marked as "i" and "ii") in the center of MTB-1; type B, a bright spot in the middle of three at the corner of MTB-1 and along the MTB-2. While mapping above (φ+ = 120 meV) or below (φ- = -72 meV) the gap, only big merged spots can be resolved. We find that the modulation of occupied (φ-) and unoccupied state (φ+) deviates from out-of-phase, as shown in Figure 2b and Figure S3a, which is quite different from the previous observation on $MoSe_2$,[10] but similar to the observation [11] on 4|4E MTB of $MoS_2$/graphene/Ir(111). We also observe an out-of-phase relationship on other MTBs of $MoTe_2$ (as shown in Figure S3b). These facts indicate that the phase of charge modulation at edge state below (φ-) and above (φ+) the gap is not an essential experimental parameter for the underlying mechanism of charge order. Thermal drift during measurement can be negligible in view of the aligned dashed triangle shapes in the two panels in Figure 2b. We also notice that the modulated bright spots are not centrosymmetric within the two end points in Figure 2b. The left distance is ~ 1.0 nm, while the right ~ 1.3 nm. This is inconsistent with a standing wave scheme with both ends fixed, where a centrosymmetric wave pattern is expected with zero amplitude at both ends.

Two parallel rows of resolved spots appear equivalent along the entire MTB, consistent with its inversion-symmetry. The distance between two equivalent spots perpendicular to MTB is measured to be ~ 7.9±0.2 Å in Figure 2a and the width of the MTB is estimated to be ~ 1.85 nm. If the MTB is a perfect isolated 1D structure without distortion, we should observe mirror symmetry structure with same contrast along MTB in tunneling mapping of the MTB region. The MTB as lattice defects would introduce local lattice strain and induce extra potential interaction on the conducting electrons, then the charge density or lattice modulate accordingly.

The resolved spots within one $3a$ period ($3a$ =10.56 Å) help us to locate the coordinate along the MTB direction quite well. Along the perpendicular direction, we have not observed any obvious displacements. Line profiles in Figure 2c show the electron density modulation with a marked position in Figure 2a. In type A structure, brighter spot "i" has an average distance of 3.27±0.07 Å with spot "ii", smaller than the regular lattice space $a$= 3.52 Å. Such displacement of ~25 pm along with the MTB corresponds to lattice distortion in 1D MTB, also called Peierls-type lattice distortion. As expected, we should observe another spot within one 3$a$ period, but it is too weak to be detected here. In type B structure at the corner of MTB-1 and MTB-2, all three spots are well resolved in Figure 2a, which enable us to measure the distortion for individual spots (Figure 2d). The distance between spot "1" and spot "2" is 3.30±0.10 Å in type B structure, while the distance between spot "1" and spot "0" is 3.14±0.07 Å. Along the MTB-1, the relative brightness of spots vary with coordination, suggesting a non-uniform lattice distortion. However, we find the gap size keeps the same along the entire MTB, indicating that the formation of the gap belongs to collective electronic behavior.

We have summarized the observation in Figure 2d and Figure 2e. We observe electron modulation is accompanied by Peierls-type lattice distortion. The modulated electron density ρ can be expressed as $\rho(r) = \rho_{avg} + \rho_0 \cdot \cos(q \cdot r + \phi(r))$, where **$q = 2\pi/\lambda = 2k_F$** is the wave vector, $\rho_{avg}$ the average charge density, $\rho_0$ the amplitude of modulation ordered parameter, λ the lattice period (here λ = 3$a$), and ϕ the phase of wave, reflecting the location of the charge modulation with respect to the underlying lattice. The coherence phase varies with position **$r$**. An electron gap of 2Δ opens at the edge of the folded Brillion zone ~ **$k_F$**. All the observations above fulfill the basic concept of the formation of CDW order in 1D MTB of MoTe$_2$.

To investigate how the MTB distortion is related to charge modulation and the opened CDW gap, we build the unit cell of the 2H-MoTe$_2$ with MTB defects in Figure 3a and calculate the electron structure *via* DFT using Perdew-Burke-Ernzerhof [23] functional within generalized gradient approximation (GGA) (see the Method). The atoms inside the red dotted box (~2.0 nm) are considered as MTB areas since their wavefunctions are affected by the MTB defects, which is consistent with the measured MTB's width of ~ 1.85 nm and the calculations in ref. 10. The atoms inside the blue box behave bulk-like electronically and the rest are the edge. Based on the measured distortion values and the detailed structure, starting with a fully relaxed 3*a'* (*a'* = 353.5 pm after relaxation) period unit cell, we distort Mo atoms (with adjacent Te atoms) marked as spot "0" and spot "2" with a magnitude of ± $d_1$ (± $d_2$) Å and leave the spot "1" Mo atoms and rest of the atoms fixed. More than 40 possible combinations of distortions were calculated, similar electronic structures with gaps are observed. On a fully relax unit cell, there is a finite metallic state rather than opened gap near $E_F$, as shown in Figure S5. To compare with the STM/STS measurements, Figure 3b shows the calculated band structure along the Γ-X direction of the system with a distortion of $d_1$ = 38 pm; $d_2$ = 22 pm, with the spot 0-1, spot 1-2, spot 2-0 distances are 315.5 pm, 331.5 pm, and 403.5 pm, respectively. The contribution from the edge atoms is removed for the irrelevancy. Energy levels inside the bulk states are apparently seen, consistent with the metallic conductance of MTB. The highest occupied state is at X point, while the lowest unoccupied state is Γ point. We find that the lowest unoccupied state is sensitive to the amount of distortions. For smaller distortion, the lowest unoccupied state locates at X point. At X point ($k_X$ = π/3*a*) of the edge of the folded Brillouin zone, and a small gap opening around 150 meV is observed in Figure 3c. The calculated local density of states (black line), projected onto the atoms inside the red dotted box of Figure 3a, is energetically aligned with the d*I*/d*V* spectrum (red line) for comparison, and the peaks agree well with each other except for some satellite peaks on the negative energy side. The satellite peaks are probably caused by the extra potential field induced by distorted lattice within one 3*a* period.

    Figure 3d shows the band decomposition onto atoms and orbitals. The band below the gap (denoted as φ-) is mainly contributed from the spot "0" and "1" Mo atoms. The case is more complicated for the band above the gap (denoted as φ+). The distribution of density states can switch from spot "1 and 2" to spot "0 and 2" if suffering from different distortions. This should be the reason why in Figure 2b we observe nearly the same shape appearance at φ-, but two types (type A and type B) of distributions are observed at φ+ with varied local distortions. Density states at φ- are formed from

orbitals perpendicular to the surface ($d_{xz}$ and $d_{yz}$), while the density states at φ+ has a certain amount of in-plane component ($d_{xy}$, $d_z^2$, and $d_{x^2-y^2}$). Such an in-plane component may help us to resolve the atomic-resolved structure at φ+. STS simulation is done by extracting LDOS in Figure 3e, which agrees well with the experimental mapping results.

To further support the formation of CDW in 1D MTB, we conducted spatial conductance measurement along the MTB (see the black arrow in Figure 1b). Figure 4a shows the obtained color plot, which clearly reveals the presence of well-separated quantized energy levels outside the energy gap $E_{gap}$ and apparent charge modulation at different energy levels. Jolie *et al.* analyzed the quantized electron state of 4|4E MTB in MoS$_2$ within the framework of the TLL scheme and related them to charge-spin separation with different traveling speed.[11] However, there are several discrepancies compared to our observation here: (1) The space of neighboring quantized energy level is approximately equidistant to 97 meV here, as shown in Figure 4b. (2) The energy gap $E_{gap}$ does not much decrease with length of MTB (3.3 nm length with $E_{gap}$ of 159±10 meV, and 5.6 nm length with $E_{gap}$ of 152±10 meV). (3) The spatial periods of charge modulation at quantized energy levels are not the same but decreased with higher number of energy level, as shown in Figure 4c. (4) The DOS peak energies are rather uniform along the entire MTB, even at the corner position. Our observations agree well with a calculated spectrum of 4|4E MTB in ref. 11 within the CDW scheme, which further demonstrate that the charge modulation in 1D MTB of monolayer MoTe$_2$ can be explained by the formation of CDW, rather than the TLL.

In Figure 4a, we find that along the entire MTB-1 the CDW gap size is nearly the same, even when there is non-uniform stress distribution. It excellently indicates the CDW is a collective electron behavior. However, it varies a lot on other MTBs with different lattice distortions. Figure 5a shows the conductance tunneling spectra on MTB-1 and MTB-2 with the same length but different surroundings. A much smaller CDW gap of ~96 meV with weaker peak features is detected on MTB-2 than that of ~152 meV on MTB-1. It is consistent with the DFT calculation in Figure 5b: the CDW gap increases linearly with the amount of lattice distortion. We notice that in Figure 5a the peak at φ- nearly is unchanged, while the peak at φ+ is shifted a lot, also agreeing well with the theory calculation in Figure 5c. Quantized energy levels with a distance of 73 meV are observed on MTB-2 *via* the line spectra. We find the quantized energy level separation positively correlates with the size of the CDW gap and lattice distortion, suggesting the electron-lattice interaction is important for the formation of CDW here.

Additionally, further information about the critical experimental parameters associated with the 1D CDW need to be discussed. (1) The observed CDW is a collective electron behavior with the same magnitude of the energy gap and the same charge modulation periodicity along the 1D chains. The strength of coherent peaks varies with locations but have certain periodicity. (2) The energy gap of the CDW state varies with distortions on different chains, while it is not sensitive to local distortion within one chain, which only induce localized varied charge density states. (3) The modulations at occupied (φ-) and unoccupied (φ+) states of the distorted MTB can deviate from out-of-phase, even in-phase. The phase relationship much depends on the projected density states onto atoms. Detailed DFT calculations (see Figure S4) show that the projected density states onto different atoms change when suffering from different or non-uniform distortions, which is also consistent with our STS mapping observations in Figure 2b and Figure S3.

**CONCLUSIONS**

In summary, by using constant-height scanning tunneling spectroscopy mapping, we achieve the detailed charge distribution within one charge modulation period of 1D MTB in monolayer $MoTe_2$. The coexisting apparent atomic-resolved periodic Peierls lattice distortions and U-shaped energy gap clearly demonstrate a Peierls-type CDW. The uniform gap with non-uniform lattice distortion validates the collective 1D CDW behavior. The DFT band structure from theoretical calculations based on measured lattice distortion agrees well with the observations and formation of CDW order. The same charge modulation period under varied mapping energies and non-centrosymmetric distribution in the whole MTB indicate the absence of quantum well states and standing waves along the MTB. We also discover the equidistant quantized peaks outside the energy gap, as well as varied periodicity along the MTB. These observations further confirm the Peierls-type CDW state in MTB of monolayer MoTe$_2$.

**METHODS**

**MBE growth**: High-quality monolayer MoTe$_2$ was grown on BLG/SiC(0001) in a commercial SPECS UHV MBE system. The structural quality and the coverage of the monolayer MoTe$_2$ samples were characterized by *in situ* reflected high-energy electron diffraction (RHEED), room temperature STM/STS, X-ray photoelectron spectroscopy (XPS) and *ex situ* Raman spectroscopy.

**LT-STM measurement**: To protect the film from contamination and oxidation during transport to the UHV LT-STM chamber, a Te capping layer with a thickness of ~10 nm was deposited on the sample surface after growth. For subsequent LT-STM measurement, the Te capping layer was removed by annealing the sample to ~600 K in the LT-STM system for 30 min. LT-STM imaging and STS measurements were performed at T=77 K in a commercial UNISOKU UHV system (USM 1300 vector magnet). STS differential conductance (d$I$/d$V$) point and line spectra were measured using standard lock-in techniques (f = 973 Hz, V$_{rms.}$ = 10 mV, T= 77 K). For d$I$/d$V$ spectra measurement, the STM Pt/Ir tip was calibrated on Au (111) surface. STM/STS data are analyzed using WSxM software.

**DFT calculation methods**

First principle calculations are performed using the Vienna Ab-initio Simulation Package (VASP),[24] which have implemented a plane-wave method with projector augmented-wave pseudopotentials, with the Perdew-Burke-Ernzerhof (PBE) functional [23] within Generalized Gradient Approximation (GGA). The cutoff energy of 400 eV is chosen for all calculations. Monkhorst-Pack k-point grid sampling [25] is used for the Brillouin zone integrations, 1×7×1 for the one-unit cell relaxation, 1×9×1 for the 3$a$ supercell total energy calculations and 3×59×3 for the electronic structure calculations. 8 Å of vacuum on each side of the MTB along perpendicular directions are added due to the imposing of periodic boundary conditions. The PBE relaxed lattice constant is 3.535 Å with the criteria of Hellmann–Feynman force on each atom less than 0.04 eV/Å, agree well with the experimental value (3.52 Å) acquired in the STM measurements. 101 k-points are used to sample the band structure along the Γ-X direction.

The band structure of the fully relax unit cell with "3$a$" period is also calculated, as shown in Figure S5. According to the calculated band structure along the Γ-X direction and projected LDOS, there is a metallic state rather than opened gap near the Fermi level, different from the observed gapped electronic state on 1D MTB here. It shows that the distorted structure agrees well with the experimental data.

**Constant-height STS mapping methods**

There are two different constant-height STS mapping methods in the experiment. The first one, which is called the normal method here, is that after scanning the STM image under certain condition (such as $V_s$ = -1.0 V, $I_t$ = 100 pA), we switch off the feedback to keep the tip's height constant, set the bias (or mapping energy) we needed and then do the mapping. The second one, named the improved method, is that we set the bias when the feedback is on, and then switch off the feedback to carry out the mapping. Compared with the normal method used in Figure 1b in the main text, we can locate the tip closer to the surface by using the second method in Figure 2a, which finally helps us to achieve atomic-resolved structures.

**ASSOCIATED CONTENT**

**Supporting Information**

The Supporting Information is available free of charge on the ACS Publications website at DOI: XXX.

Additional figures pertaining to STM observations of (1×1) MoTe$_2$ domain, mapping of 1D MTB under various energies, line spectroscopy, mapping with the constant-height mode, LDOS under different distortions and band structure of the fully relaxed unit cell (PDF).

**AUTHOR INFORMATION**

**Corresponding Authors:**

\* E-mail: fsli2015@sinano.ac.cn

\* E-mail: adingsun2014@sinano.ac.cn11

* E-mail: wangguang@nudt.edu.cn

**Author Contributions**

⊥L.W., Y.W., and Y.Y. contributed equally.

**Notes** The authors declare that they have no competing interests.

**ACKNOWLEDGMENTS**

We thank Y. S. Fu and F. Liu for helpful discussions. G.W. acknowledges financial support from the National Natural Science Foundation of China (NSFC) (No. 11574395) and Natural Science Foundation of Hunan Province, the research project of National University of Defense Technology (No. ZK18-03-38) and the Open Research Fund Program of the State Key Laboratory of Low-Dimensional Quantum Physics (No. KF201904). F. L. and A. D. acknowledge support from NSFC (No. 11604366 and No. 11634007) and the Natural Science Foundation of Jiangsu Province (No. BK 20160397). F. L. acknowledges support from the Youth Innovation Promotion Association of Chinese Academy of Sciences (2017370).

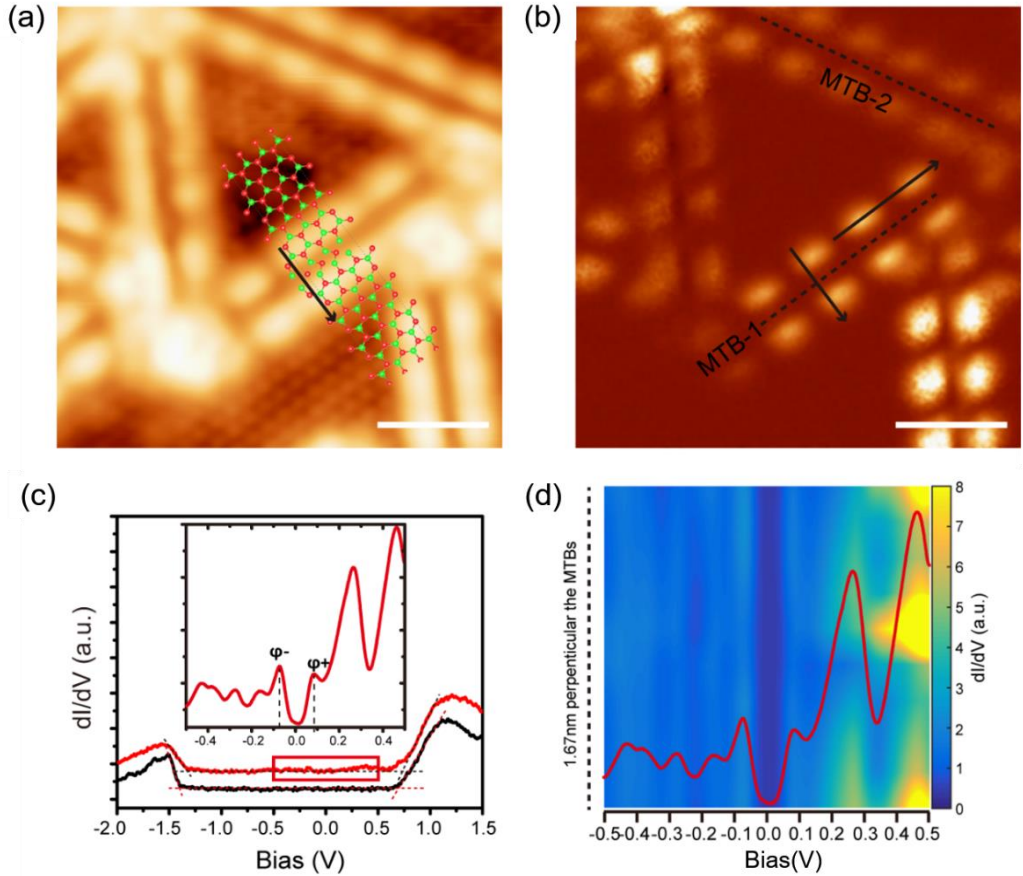

Figure 1. Morphology and electronic structure of 1D MTB in monolayer MoTe$_2$ on BLG/SiC(0001) substrate (color online). (a) STM image of monolayer MoTe$_2$ on BLG/SiC(0001) ($V_s$ = -1.0 V, $I_t$ = 100 pA, scale bar: 2 nm), showing triangle-shaped domain boundaries. The ball model gives a detailed structure based on the previous high-resolution AFM image.[10] (b) Representative d$I$/d$V$ normal constant-height conductance mapping recorded at energy of 73 mV (starting condition: $V_s$ = -1.0 V, $I_t$ = 100 pA). Apparent charge modulations with 3$a$ periodicity could be identified along triangle edges (MTB-1, MTB-2 and so on). Black arrows show the location of line spectroscopy taken along or perpendicular to the MTB. Scale bar: 2 nm. (c) Typical d$I$/d$V$ spectra acquired on pristine monolayer MoTe$_2$/BLG (black line) and on an MTB (red line), An enlarged view on MTB in the inset shows clear gapped structure around the Fermi level. We label the negative (positive) peak as φ- (φ+). (d) Color plot of spatially resolved d$I$/d$V$ tunneling spectra taken perpendicular to MTB-1 (position of black arrow perpendicular to MTB-1 in Figure 1a and Figure 1b), showing rather uniform electric structures, consistent with its 1D nature.

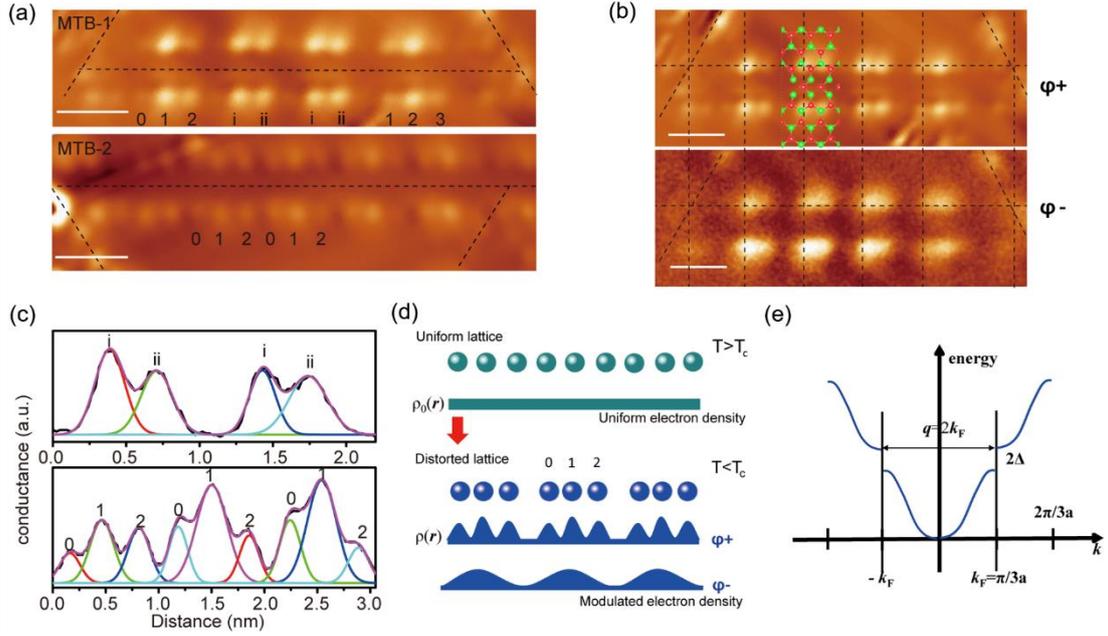

Figure 2. High-resolution d$I$/d$V$ mapping and lattice distortion of 1D MTB in monolayer MoTe$_2$ (color online). (a) Tunneling conductance maps of MTB-1 and MTB-2 using improved constant-height mode (starting and mapping condition: $E$ = 80 meV, $I_t$ = 100 pA, scale bar: 1 nm), showing well-resolved structures inside one charge modulation period of 3$a$. We label the bright spots accordingly. (b) Representative d$I$/d$V$ constant-height conductance maps of MTB-1 recorded at energies corresponding to peak states (φ+ at 80 meV and φ- at -72 meV) of the opened gap in Figure 1c. The dashed lines are guided to the eyes. (c) Line profiles of marked spots in the MTB-1 and MTB-2. The curves are fitted with the Gaussian function, which facilitates the measurement of peak distance. (d) Schematic diagram of the formation of Peierls-type CDW below CDW transition temperature (T$_c$) with distorted lattice and the corresponding charge modulation. (e) Schematic diagram of the band structure of Peierls-type CDW, showing the formation of the CDW gap (2$\Delta$) at the edge of the folded Brillion zone.

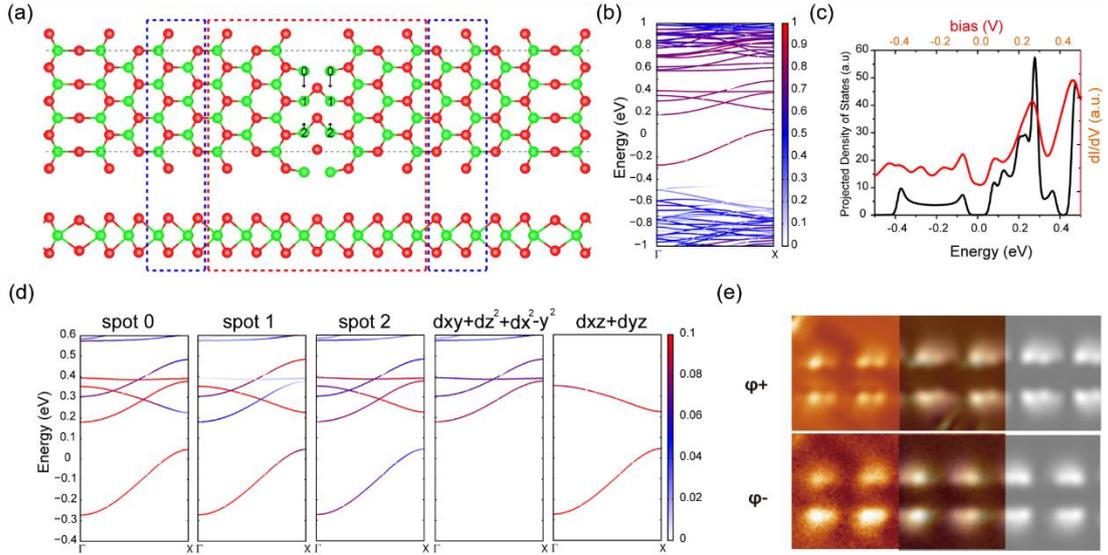

Figure 3. Calculated electron structure of 1D MTB in monolayer MoTe$_2$ (color online). (a) Ball model of 1D MTB (Red: Te, Green: Mo). The red dashed box marks the MTB area, while two blue dashed boxes show a bulk-like electron state. The position of Mo atom labeled "1" at MTB area is fixed during the calculation, while "0" and "2" atoms distorted based on the measured distance between position "1". (b) Calculated band structure along Γ-X direction, showing metallic state within the normal bandgap. A gap is observed at the X point. (c) The DOS (black line) projected onto the atoms inside the red dotted boxes of Figure 3a aligns well with the measured STS (red line), except some satellite peaks at negative bias. (d) Projected band structure. The left two panels show the "1" and "2" atoms contribute to positive peak φ+, while "0" and "1" atoms to negative peak φ-. The right two panels show projected band structures onto Mo 4d orbitals ($d_{xy}+d_{z^2}+d_{x^2-y^2}$ and $d_{xz}+d_{yz}$), suggesting the electronic state at φ- is due to vertical component "$d_{xz}+d_{yz}$", while that at φ+ can have in-plane component when suffering from large distortions. The color range indicates the electron occupation. (e) Simulated LDOS mapping at φ- band and the φ+ agree well with experimental STS mapping.

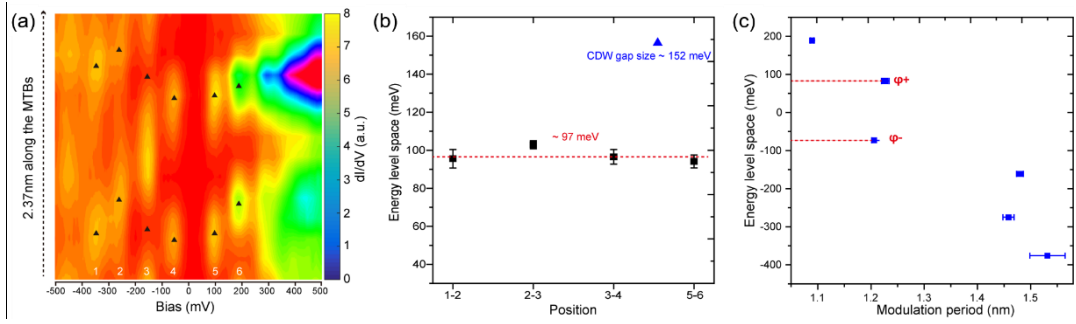

Figure 4. Electronic structure along the 1D MTB in monolayer MoTe$_2$ (color online). (a) The color plot of d$I$/d$V$ line profile along MTB-1 in Figure 1b with a length of 2.37 nm, showing spatially electron distribution along 1D MTB. Satellite peaks (from "1" to "6") in STS spectra show clear periodical modulations along the 1D MTB, which are marked by small filled triangles. (b) The energy differences between neighboring quantized energy level, showing equidistant space of 97meV, much smaller than the CDW gap of 152 meV. (c) The spatial periods of charge modulation of quantized energy levels shown in Figure 4a, indicating larger charge modulation periodicity at φ- than that at φ+.

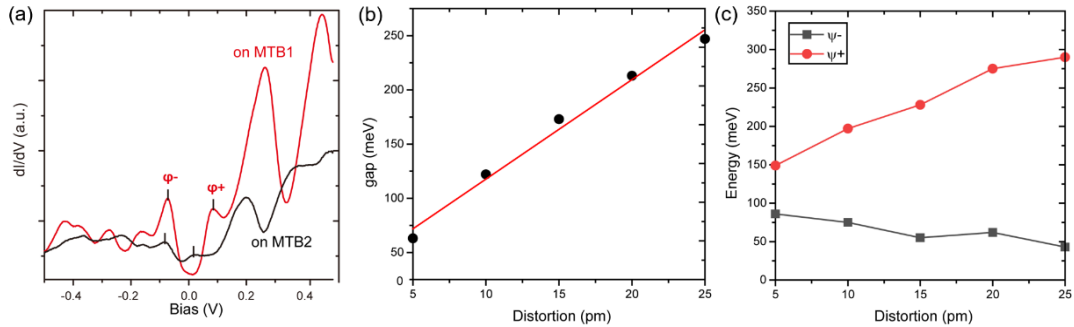

Figure 5. The relationship between CDW gap size and lattice distortion (color online). (a) d$I$/d$V$ spectra along the MTB-1 (red curve) and MTB-2 (black curve), showing the CDW gap in MTB-2 is much suppressed. The position of the φ- band is nearly the same, while φ+ band varies a lot. (b) DFT calculations show the CDW gap nearly linearly increased with the amount of distortion. (c) The calculated energy level of φ- (black curve) and φ+ (red curve) at X point as a function of the amount of distortion, showing the energy position of φ+ varies a lot under different distortions, consistent with the experimental observation.